\documentclass[%
 reprint,
superscriptaddress,
 amsmath,amssymb,
 aps,
prl,
]{revtex4-2}

\usepackage{graphicx}
\usepackage{dcolumn}
\usepackage{bm}
\usepackage{ulem}
\usepackage{epstopdf}
\usepackage[T1]{fontenc}
 \usepackage[usenames,dvipsnames]{pstricks}
 \usepackage{epsfig}
 \usepackage{pst-grad} 
 \usepackage{pst-plot} 
 \usepackage[space]{grffile} 
 \usepackage{etoolbox} 
 \makeatletter 
 \patchcmd\Gread@eps{\@inputcheck#1 }{\@inputcheck"#1"\relax}{}{}
 \makeatother
\usepackage{booktabs}
\usepackage{boxhandler}
\usepackage{amsmath}
\usepackage{makecell}
\usepackage{bbold}
\usepackage{natbib}
\usepackage[whole]{bxcjkjatype}
\usepackage{tikz}
\usepackage{tabularx}
\usepackage{mathtools}
\usetikzlibrary{arrows.meta}
\usepackage{hyperref}
\hypersetup{
  colorlinks   = true,    
  urlcolor     = blue,    
  linkcolor    = blue,    
  citecolor    = blue      
}
\usepackage[caption=false]{subfig}
\usepackage{floatrow}
\usepackage[utf8]{inputenc}
\usepackage{braket}
\usepackage[cmintegrals]{newtxmath}
\usepackage{multirow}
\usepackage{enumitem}
\usepackage{pifont}

\begin{document}

\title{SpinHex: A low-crosstalk, spin-qubit architecture based on multi-electron couplers}

\author{Rubén M. Otxoa}
\email{ruben.otxoa@hitachi-eu.com}
\affiliation{Hitachi Cambridge Laboratory, J. J. Thomson Avenue, Cambridge CB3 0HE, United Kingdom.}
\author{Josu Etxezarreta Martinez}
\affiliation{Department of Basic Sciences, Tecnun - University of Navarra, 20018 San Sebastian, Spain.}
\affiliation{Cavendish Laboratory, Department of Physics, University of Cambridge, Cambridge CB3 0HE, UK.}
\author{Paul Schnabl}
\affiliation{Department of Basic Sciences, Tecnun - University of Navarra, 20018 San Sebastian, Spain.}
\author{Normann Mertig}
\affiliation{Hitachi Cambridge Laboratory, J. J. Thomson Avenue, Cambridge CB3 0HE, United Kingdom.}
\author{Charles Smith}
\affiliation{Hitachi Cambridge Laboratory, J. J. Thomson Avenue, Cambridge CB3 0HE, United Kingdom.}
\author{Frederico Martins}
\affiliation{Hitachi Cambridge Laboratory, J. J. Thomson Avenue, Cambridge CB3 0HE, United Kingdom.}

\begin{abstract}

Semiconductor spin qubits are an attractive quantum computing platform that offers long qubit coherence times and compatibility with existing semiconductor fabrication technology for scale up.
Here, we propose a spin-qubit architecture based on spinless multielectron quantum dots that act as low-crosstalk couplers between a two-dimensional arrangement of spin-qubits in a hexagonal lattice.
The multielectron couplers are controlled by voltage signals, which mediate fast Heisenberg exchange and thus enable coherent multi-qubit operations.
For the proposed architecture, we discuss the implementation of the rotated XZZX surface code and numerically study its performance for a circuit-level noise model. We predict a threshold of $0.18\%$ for the error rate of the entangling gates.
We further evaluate the scalability of the proposed architecture and predict the need for $4480$ physical qubits per logical qubit with logical error rates of $10^{-12}$ considering entangling gate fidelities of $99.99\%$, resulting in a chip size of $2.6$cm$^2$ to host $10,000$ logical qubits.

\end{abstract}

\maketitle

\section{Introduction}

Fault-tolerant quantum computers enable calculations that are impractical for classical supercomputers \cite{flannigan2022propagation, beverland2022assessing, lee2023evaluating, hoefler2023disentangling}. 
However, building a fault-tolerant quantum computer is difficult.
For example, it requires managing hardware errors that limit the complexity of addressable computations \cite{shor1996fault, shor1995scheme, steane1996error, cai2023quantum, temme2017error, VQELimitationsI, VQELimitations, dalton2022variational}.
Thus, error correction is central for realizing large-scale, fault-tolerant quantum computers.

Surface codes \cite{fowlerSC,xzzx,decoders} are a promising approach to realizing error-corrected quantum hardware.
They allow for suppressing errors by combining many physical qubits that operate as a single logical qubit.
The feasibility of this error-suppression concept has recently been demonstrated on superconducting and neutral atom hardware \cite{surfacecodeETH, googleSurf, google24, neutralQEC} for a single logical qubit.
However, executing complex quantum algorithms requires many (logical) qubits with long coherence times \cite{dalzell2023quantum}, which remains a significant challenge for any quantum hardware.

Semiconductor spin qubits are one potential platform for realizing error-corrected quantum hardware \cite{PhysRevA.57.120, RevModPhys.85.961, RevModPhys.95.025003}.
Their benefits include small footprints, long coherence times \cite{Stano2022, Tanttu2024}, high gate fidelity \cite{Yoneda2018, Xue2022, Noiri2022, doi:10.1126/sciadv.abn5130, Veldhorst2015Oct, Huang2024}, operation at 1K \cite{Huang2024}, and the potential for scale-up using semiconductor fabrication technology \cite{Zwerver2022, 10185272, 9371956, elsayed2022low, Neyens2024}.
Although experimental verifications of error correction protocols have just started \cite{takeda2022quantum}, several theoretical works propose semirealistic architectures \cite{DzurakArchitecture2017, vandersypen2017interfacing, Li2018Crossbar, boter2022spiderweb, Crawford2023, kunne2023spinbus, Cai2023, Siegel2024_2xN, siegel2025snakesplanemobilelow} that allow for implementing surface codes with spin quibts.
Unfortunately, many proposals suffer from high wiring densities and partially rely on long-range coupling methods \cite{nichol2017, shulman2012, borjans2020resonant, van2018microwave, dijkema2023two}, which have not yet achieved the required fidelity.
Some proposals \cite{boter2022spiderweb, kunne2023spinbus}, on the other hand,  significantly reduce the wiring density, by shuttling spin qubits from occupied to vacant quantum dots with high fidelity.
The feasibility of this approach is under active investigation, mainly in one-dimensional (1D) arrays of quantum dots  \cite{Shuttling2020, YonedaShuttling, ShuttlingTarucha2022, SiGeShuttle2023, ShuttleBlueprint2023, QuBus2024, ConveyerMode2024, QuBus2024, ConveyerMode2024, riggelen2024coherent, desmet2024highfidelitysinglespinshuttlingsilicon}.
Meanwhile, experimental investigations of two-dimensional (2D) spin-qubit devices are still rare \cite{Hendrickx2021, Lim2024, Unseld2023, riggelen2024coherent, Borsoi2024, John2024}.
Thus, we anticipate challenges with the two-dimensional integration necessary for implementing surface codes.
In particular, we expect issues due to (capacitive) crosstalk in the couplers \cite{boter2022spiderweb, kunne2023spinbus}, which connect 1D arrays of quantum dots in the necessary second dimension.

In this paper, we propose SpinHex -- a semirealistic device design for operating surface codes with spin qubits at low crosstalk.
SpinHex provides a 2D arrangement of spin qubits in a hexagonal lattice.
Its core unit is a spinless multi-electron quantum dot, which mediates voltage-controlled Heisenberg exchange \cite{mehl2014two, srinivasa2015tunable, Malinowski2019fast} between up to three spin qubits in 2D.
Such a multi-electron coupler (MEC) provides a large separation between the coupled spin qubits, thus significantly lowering capacitive crosstalk.
To evaluate the feasibility of SpinHex, we discuss the implementation of the rotated XZZX surface code \cite{xzzx} and numerically study its performance under device-specific circuit-level noise.
In particular, we consider the errors introduced by swapping spin-qubits across MECs and report a threshold of 0.18\% on the error rate of the entangling gates.
We also compute the number of physical qubits required to reach the \textit{MegaQuop}, \textit{GigaQuop}, and \textit{TeraQuop} regimes, where a single logical qubit can perform a million, a billion, or a trillion error-free quantum operations, respectively. 
At error rates of 0.01\% we show that approximately 1,000 physical qubits are required to reach the MegaQuop regime. Meanwhile, a footprint of approximately 4480 qubits is required to reach the TeraQuop regime.
Finally, we estimate that a chip operating 10,000 logical qubits in the TeraQuop regime would result in a chip of around $2.6\mathrm{cm}^2$. 

\begin{figure*}[]
\centering
\includegraphics[width=0.9\textwidth]{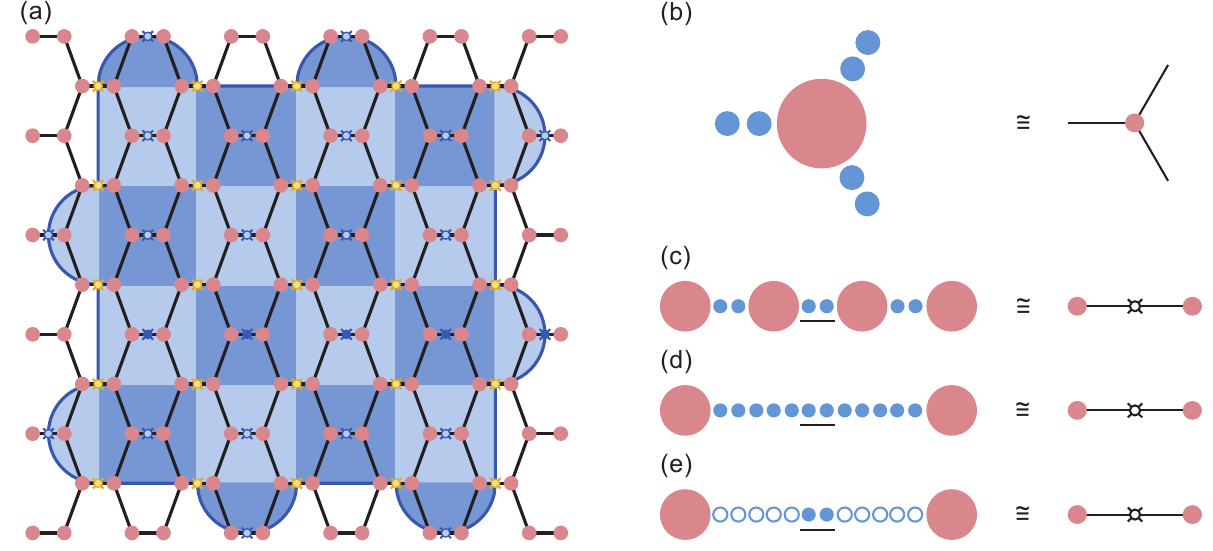}
\caption{
(a) SpinHex hardware graph composed of MECs (red discs) and 1D quantum dot arrays (black lines). Cirles (yellow and blue) represent the data and measurement qubits of a logical qubit (blue background tiles).
(b) MEC coupling the occupied quantum dots (blue disc) of three 1D quantum dot arrays.
(c-e) Three possible realizations of a 1D quantum dot array. Blue circles represent empty quantum dots. The underlined quantum dots host the surface code qubit.
}
\label{fig:HardwareGraph}
\end{figure*}

\section{Results}

\subsection{SpinHex hardware graph}

The hardware graph of SpinHex is depicted in Fig.~\ref{fig:HardwareGraph}.
Similar to~\cite{kunne2023spinbus}, SpinHex connects 1D arrays of quantum dots via three-way junctions to form a hexagonal lattice of data and measurement qubits, see Fig.~\ref{fig:HardwareGraph}(a).
Instead of T-junctions \cite{kunne2023spinbus}, we propose using spinless multi-electron (or hole) quantum dots \cite{mehl2014two, srinivasa2015tunable, Malinowski2019fast} to couple the terminal ports of three 1D quantum dot arrays in 2D, see Fig.~\ref{fig:HardwareGraph}(b).
Such multi-electron (or hole) couplers (MECs) provide a large separation between the terminal quantum dots of the coupled 1D quantum dot arrays, resulting in low capacitive crosstalk.
Finally, we envision three possible realizations of the 1D quantum dor arrays, as depicted in Figs.~\ref{fig:HardwareGraph}(c-e):\

In the first realization, Fig.~\ref{fig:HardwareGraph}(e), the 1D array consists of empty quantum dots.
Two occupied quantum dots host a compute and reference spin-qubit for readout at the array's center.
The spin-qubit is shuttled along the 1D array to SWAP its information with other 1D arrays across the MECs at the terminal.
This realization is similar to replacing T-junctions in Ref.~\cite{kunne2023spinbus} with three-way MECs.
Due to the high fidelity of shuttling operations, this is likely an interesting realization.

In the second realization, Fig.~\ref{fig:HardwareGraph}(d), the quantum dots of the 1D array are all occupied by one spin-qubit.
Two quantum dots host a compute and reference spin-qubit for readout at the array's center.
In this realization, the spin-qubit is swapped to the terminal ports of the array, to exchange information with other 1D arrays across an MECs at the terminal.
This realization is reminscent of Ref.~\cite{Crawford2023}.
Due to the low fidelity of SWAP operations, this is likely an inferior realization.

Finally, in this paper we focus on a third realization, Fig.~\ref{fig:HardwareGraph}(c), where the 1D array consists of double quantum dots connected via MECs.
Due to the large size of the MECs the double quantum dots are expected to have low capacitive crosstalk.
This configuration is expected to simplify device tuning and facilitate low-crosstalk.
Two quantum dots at the array's center host a compute and reference spin-qubit for readout.
In this realization, the spin-qubit information is swapped within the double dots as well as across the MECs to reach the terminal ports of a 1D array, from where it swaps information with other 1D arrays across the MECs.

\subsection{SpinHex device designs}

\begin{figure*}[!ht]
\center
\includegraphics[width=0.9\textwidth]{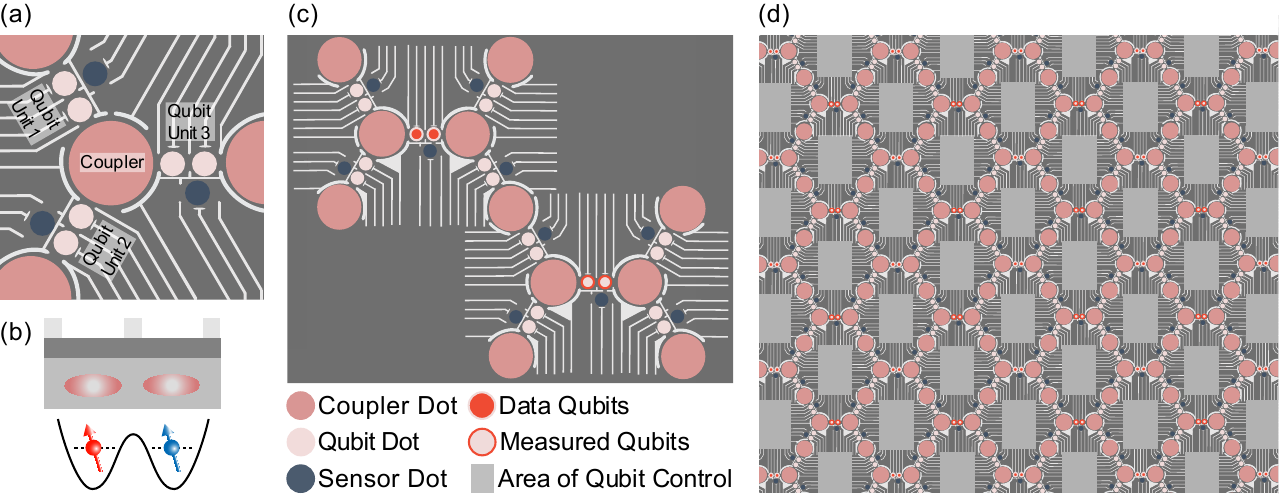}
\caption{SpinHex device design.
(a) Three way coupler:\ Metal gates (light gray) on an oxide layer (dark gray) define a central coupler (red disc) and three qubit units. Each qubit unit has a charge sensor (blue disk) and a double quantum dot (pink disks), which host a compute and a reference spin-qubit for readout, respectively.
(b) Cross section of a qubit unit:\ Metal gates (light gray) on an oxide (dark gray) on a semiconductor layer (medium gray) define a confining potential (black lines) to trap charges (pink clouds), which carry spin qubits (arrows).
(c) SpinHex unit cell:\ Accumulation gates on oxide define several couplers and double-quantum dot qubit units. Two units are marked as data and measurement qubit, respectively.
(d) Zoom on SpinHex device comprising of $4.5 \times 4.5$ unit cells. Metal gates terminate in connection islands (gray regions).}
\label{fig:DeviceDesign}
\end{figure*}

A semirealistic SpinHex device is shown in Fig.~\ref{fig:DeviceDesign}.
For simplicity, we focus on a device stack of three layers; see Fig.~\ref{fig:DeviceDesign}(b):\
A semiconductor material hosting a two-dimensional gas of charge carriers at the bottom;
An oxide layer to provide electrical insulation in the middle;
And a layer of metal gates, which create a confining potential in the two-dimensional gas of charge carriers.

The key element of SpinHex is a three way coupler, depicted in Fig.~\ref{fig:DeviceDesign}(a).
The metal gates are arranged to form a large quantum dot at the center, which is surrounded by three qubit units.
Each qubit unit has a charge sensor and a double quantum dot.
Each quantum dot is occupied with a single spin qubit.
One of the spin qubits serves as the compute qubit, while the other qubit serves as the reference qubit for readout.
The large quantum dot at the center is occupied by 50-100 charges, which form a spinless quantum dot.
It acts as a coupler that mediates Heisenberg exchange between the adjacent spins of the three qubit units.

To make SpinHex modular, we define a unit cell as depicted in Fig.~\ref{fig:DeviceDesign}(c).
In this unit cell, we use 4 three-way couplers and expand its qubit units into 1D arrays, in the spirit of Fig.~\ref{fig:HardwareGraph}(c).
To define a specific architecture, we use horizontal 1D arrays with $N_x\ge2$ couplers and diagonally oriented 1D arrays with $N_y\ge3$ couplers.
Fig.~\ref{fig:DeviceDesign}(c) shows an example where $(N_x, N_y)=(2,3)$.
Each unit cell has one data qubit and one measurement qubit along the horizontal quantum dot arrays.
To form the complete SpinHex architecture, we repeat the unit cell, as shown in Fig.~\ref{fig:DeviceDesign}(d) for a unit cell, where $(N_x, N_y)=(2,3)$.
The metal gates connect to rectangular (gray-shaded) regions, where they need to be wired to classical control electronics. We call this region the connection islands.

We now discuss several aspects of SpinHex:\
(i) SpinHex should be a valid proposal for a range of semiconductor substrates (e.g. SiMOS, Si/SiGe, Ge/SiGe, GaAs, etc.) and it should also be valid both for electron and hole spin qubits.
However, for simplicity, the device design focuses on depletion gates, commonly used in Ge/SiGe or GaAs devices. 
Similar designs in SiMOS, Si/SiGe should be possible but would require modifications to turn the disk-like structures (coupler, charge sensors, and qubit dots) into accumulation gates, e.g., by connecting them to a corresponding plunger.
(ii) While not explicitly specified in our design, SpinHex may require additional elements. For SiMOS or Si/SiGe devices operated with electron spins these include antennas or micromagnets needed to realize single spin rotations.
(iii) The coupler size is large, and thus ensures low crosstalk between spins from disjoint qubit units. However, increasing the coupler size too much will increase its level density, which may result in higher thermal noise. Choosing the optimal coupler size that balances these effects remains future work.
(iv) Separating the data and measurement qubits by a large distance may reduce the correlations of their respective charge-noise environments. This is beneficial since the surface code works best when individual errors occur independently.
(v) Especially for large values of $(N_x, N_y)$ it may be possible to integrate classical electronics within the connection island. (See~\cite{boter2022spiderweb, kunne2023spinbus, Crawford2023} for similar comments.)
Given that most qubits of a unit cell require identical control signals, multiplexing controls within connection islands could be largely beneficial. 
Finally, error correction requires decoders that process the measured syndrome information quickly \cite{decoders}. Sufficiently large connection islands may be useful for integrating application-specific decoders on chip to realize ultra-low latencies.

\begin{figure*}[]
\centering
\includegraphics[width=13cm]{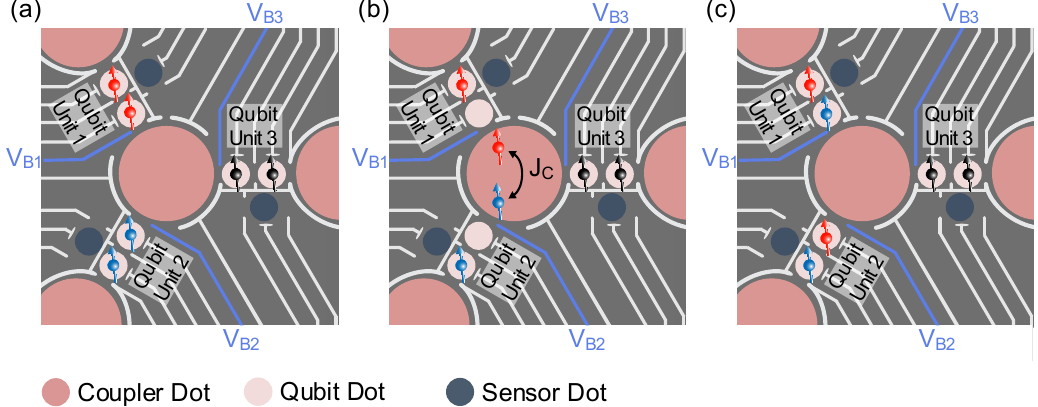}
\caption{SWAP operation across a three way coupler [details cf.~Fig.~\ref{fig:DeviceDesign}(a)]. Arrows (red, blue, and black) denote spins from qubit unit 1, 2, and 3, respectively. Blue gates $(V_{B1},V_{B2},V_{B3})$ control the barriers between the coupler and the qubit units 1, 2, and 3, respectively. (a) Initial spin-qubit configuration. (b) A SWAP operation between qubit unit 1 and 2, lowering barriers via $V_{B1}$ and $V_{B2}$, while keeping $V_{B3}$ closed. (c) Spin-qubits after the SWAP operation. 
}
\label{fig:Swap}
\end{figure*}

\subsection{Device operation}

\textbf{Device tuning:} 
To power on the device, the metal gates are set at an appropriate voltage.
The charge sensors are then supplied with charges from the reservoirs in the connection islands and the charge sensors are calibrated \cite{hanson2007spins}.
Next, the quantum dots in the qubit unit and couplers are supplied with electrons from reservoirs at the device's circumference.
Charge sensors are used to monitor charge occupation.
The couplers are occupied with a number of charges, which ensure a spinless quantum dot.
Quantum dots in the qubit unit are occupied by a single spin qubit each.
Standard tuning techniques are used to virtualize the  gates of the double dots \cite{Mills2019}.

\textbf{State preparation:} 
The spin qubits are initialized with a standard procedure \cite{Huang2024}. First, each pair of quantum dots is brought to the (0,2) charge configuration. This allows the system to relax to the singlet ground state. The gate voltages are then ramped adiabatically to reach the (1,1) charge configuration. This ramp preserves the spin state, initializing the system in the (up, down) state. One of these spins is the compute qubit and the other spin serves as a readout reference.

\textbf{Coupler SWAP:}
The key feature of SpinHex is the ability to swap spins from any pair of qubit units adjacent to the coupler.
The control sequence is similar to the 1D case, discussed in Ref.~\cite{Malinowski2019fast}.
An example of swapping the spins between the qubit units 1 and 2, while leaving the qubits in unit 3 unchanged, is given in Fig.~\ref{fig:Swap}.
Initially, the qubit system has a well-defined spin state in the quantum dots, see Fig.~\ref{fig:Swap}(a).
To execute the swap operation, the voltages on gates $V_{B1}$ and $V_{B2}$ are adjusted to lower the barriers between qubit units 1,2 and the coupler respectively. This induces a Heisenberg exchange $J_C$ \cite{mehl2014two, srinivasa2015tunable, Malinowski2019fast}, between the spins of qubit units 1 and 2, which are adjacent to the coupler, see Fig.~\ref{fig:Swap}(b). The voltage $V_{B3}$ is kept, so that the barrier between qubit unit 3 and the coupler is closed. To enable a full spin swap, the barriers are kept in this configuration for a time $t_{SWAP}$.
At the end of the SWAP operation, Fig.~\ref{fig:Swap}(c), the voltages $V_{B1}$ and $V_{B2}$ are reset to their original value. The barriers between the coupler and all qubit units are now closed.

\textbf{Qubit operations:}
Standard single-qubit rotation \cite{Stuyck2024, Yoneda2018, Hendrickx2021, Camenzind2022, Piot2022, Bassi2024} and two-qubit operations \cite{Tanttu2024, Xue2022, Noiri2022, zajac2018resonantly, Mills22, Wang2024, Geyer2024, Liles2024} can be performed in the double dot system both for electron-spin \cite{Stuyck2024, Yoneda2018, Tanttu2024, Xue2022, Noiri2022, zajac2018resonantly, Mills22} and hole-spin \cite{Hendrickx2021, Camenzind2022, Piot2022, Bassi2024, Wang2024, Geyer2024, Liles2024} qubits.
Single-qubit operations for electron spins require additional antennas to perform electron spin resonance (ESR) \cite{Stuyck2024, Huang2024} or additional micromagnets to perform electron-spin dipole resonance (EDSR) \cite{Yoneda2018}. Hole-spin qubits are controlled by voltage signals using EDSR \cite{Hendrickx2021, Camenzind2022, Piot2022, Bassi2024}.

Between the two spins of a double-quantum dot one can implement exchange driven two-qubit operations, e.g., a controlled-Z (CZ) rotation \cite{Tanttu2024, Xue2022, Noiri2022}.
These rotations are implemented by controlling the exchange between the two spins of a double-quantum dot via the voltage on the barrier between the two spins.
This technique can be applied both to electron- \cite{Tanttu2024, Xue2022, Noiri2022} or hole-spin \cite{Liles2024} qubits.
Controlled X-rotations (CX) can be achieved by adjusting both the voltage barrier and the ESR-/EDSR control signal, as e.g. implemented for electrons in \cite{Noiri2022, zajac2018resonantly} and holes in \cite{Geyer2024}. Alternatively, one can conjugate a CZ gate with a single-qubit Hadamard (H) operation.
A two-qubit operation between compute qubits of disjoint qubit units, requires swapping both qubits into the same double dot.
Such SWAP operations involve both swapping across the coupler as well as swapping spins within the compute unit. SWAP operations within a qubit unit can be composed from primitive 
single-qubit and CZ (CX) operations. Alternatively, they can be induced using Heisenberg exchange.

\textbf{Readout:}
Spin readout \cite{Oakes23, Takeda2024, Hendrickx2021, Geyer2024} is performed using, e.g., Pauli-spin blockage spin-to-charge conversion methods with current or rf-sensing via the SET.

\begin{figure*}[]
\centering
\includegraphics[width=15cm]{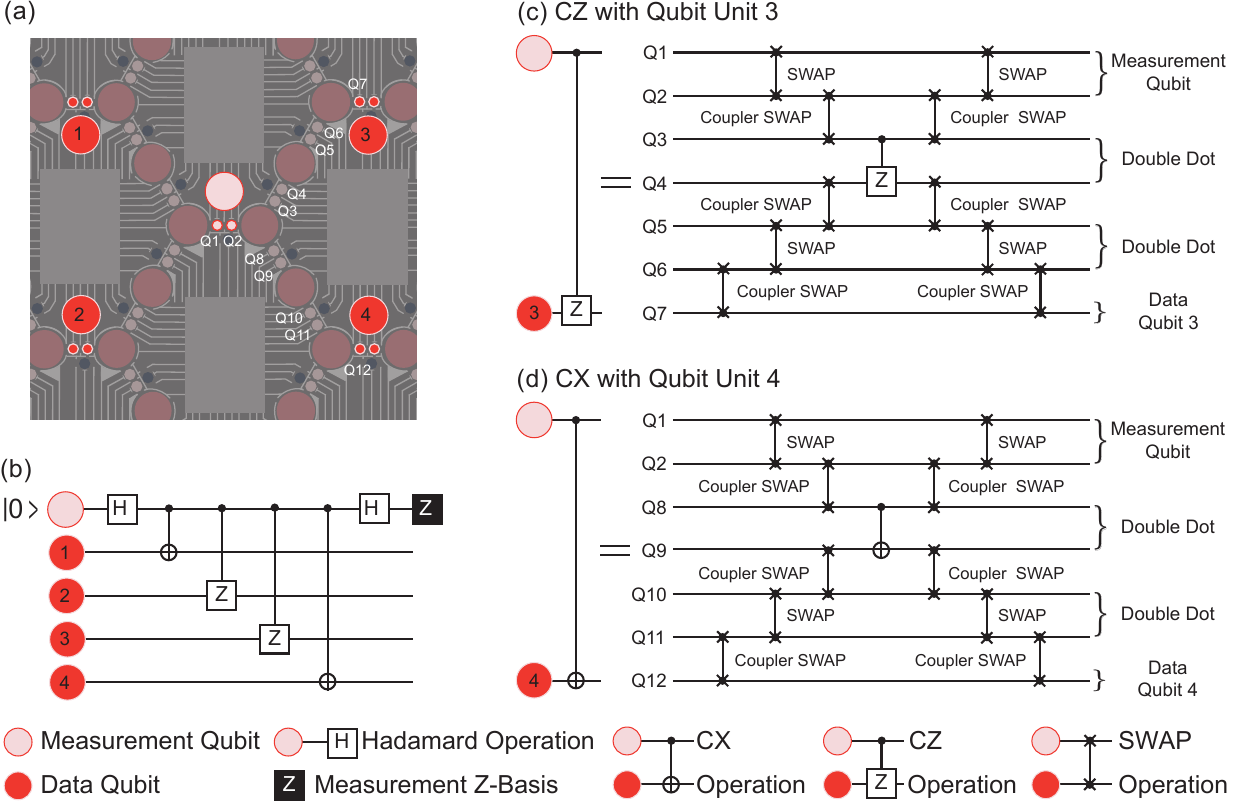}
\caption{XZZX surface code on SpinHex. (a) Section of a SpinHex device (cf.~Fig.~\ref{fig:DeviceDesign}) with a measurement and four data qubit units (white and red discs), respectively. Each qubit unit has its compute qubit at its left and its reference qubit at its right. Labels (Q1-Q12) mark qubits along the path connecting the measurement qubit unit to the data qubit units 3 and 4, respectively.
b) XZZX stabilizer circuit.
(c,d) Examples illustrating the implementation of two-qubit stabilizer operations (CZ,CX) on SpinHex by means of SWAP operations. (c) CZ between the compute qubit of the measurement qubit unit and the compute qubit of data qubit unit 3. (d) CX between the compute qubit of the measurement qubit unit and the compute qubit of the data qubit unit 4.
}
\label{fig:circuit}
\end{figure*}

\subsection{XZZX surface code on SpinHex}

A SpinHex device consisting of a $(d+1) \times (d+1)$ square lattice of unit cells can always host a rotated surface code of distance $d$. 
See Fig.~\ref{fig:HardwareGraph}(a) for an example where the surface code has distance $d=5$.
Note that such a surface code has $d^2$ data qubits and $d^2-1$ measurement qubits.
We focus on the XZZX code \cite{xzzx}, due to its consistent performance in the presence of biased noise that often appears in spin qubits.

The implementation of the XZZX code on SpinHex is illustrated in Fig.~\ref{fig:circuit} for a device with a $(N_x,N_y)=(2,3)$ unit cell.
To implement the XZZX code one performs the stabilizer circuit of Fig.~\ref{fig:circuit}(b). This stabilizer circuit must be applied to each of the $d^2-1$ measurement qubits. For each measurement qubit, implementing its stabilizer circuit requires acting on its four adjacent data qubits. Fig.~\ref{fig:circuit}(a) illustrates the typical allocation of a measurement qubit and its four adjacent data qubits on a SpinHex device.
Note that we implicitly assume that each qubit unit has its compute qubit on its left and its reference qubit on its right.
We implement the initialization, the Hadamard operation, and the readout of the stabilizer circuit on SpinHex as discussed in the previous section.
The two-qubit operations (CZ and CX) of the stabilizer circuit require additional SWAP operations, as illustrated in Fig.~\ref{fig:circuit}(c,d).
Note that for a $(N_x,N_y)$ unit cell, the implementation of a CZ (or CX) operation requires $[4(N_x+N_y)-10]$ additional SWAP operations.
(To bring the data and the measurement qubit into the same qubit unit, we require $N_x+N_y-2$ coupler SWAPs and $N_x+N_y-3$ SWAP operations within qubit units.
The same amount of SWAPs is required to return the data and the measurement qubit to their original locations.)
\floatsetup[figure]{style=plain,subcapbesideposition=top}

\begin{figure*}[!ht]
\centering
\sidesubfloat[]{\includegraphics[width=7.9cm]{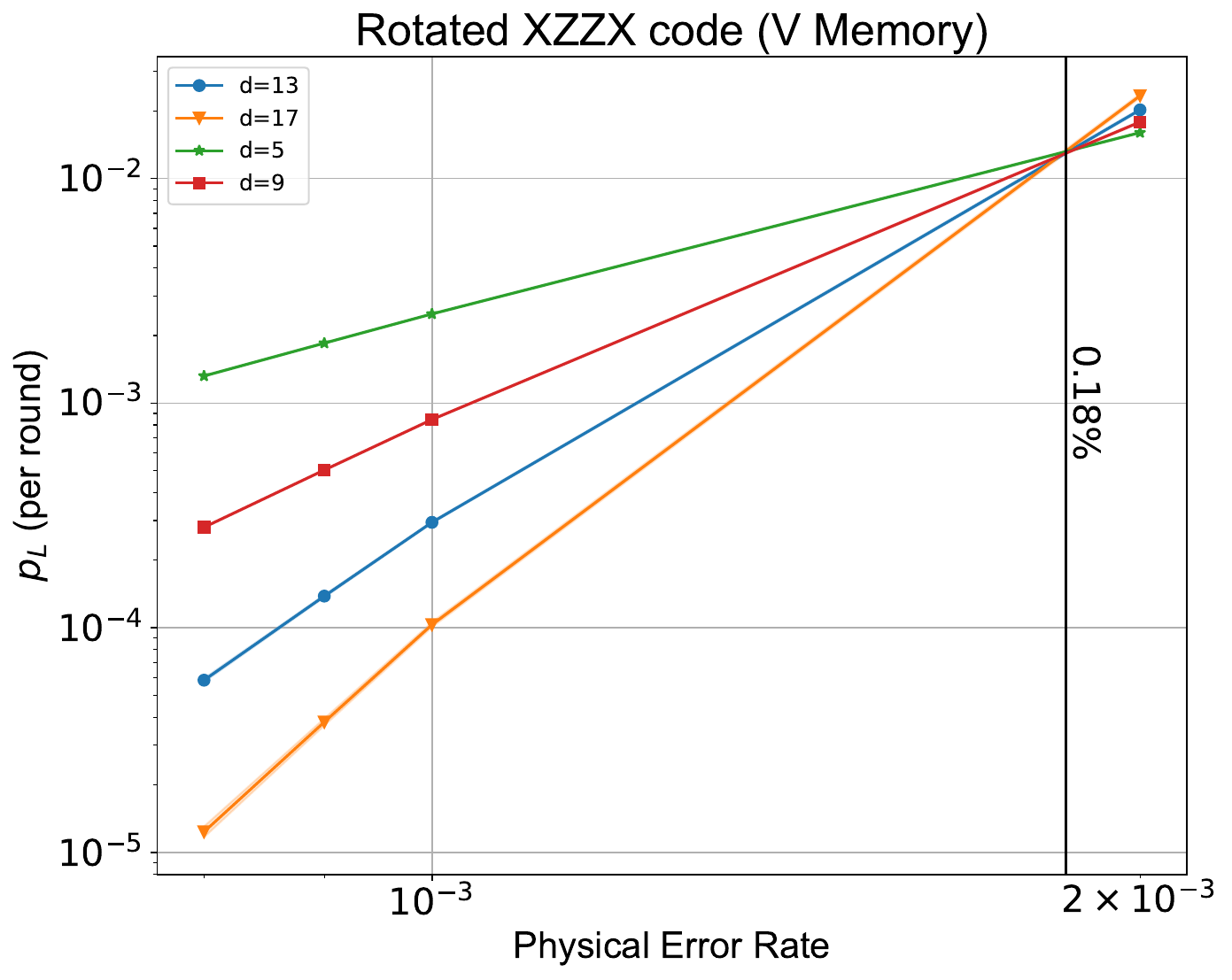}\label{fig:wat1}}\quad%
\sidesubfloat[]{\includegraphics[width=7.9cm]{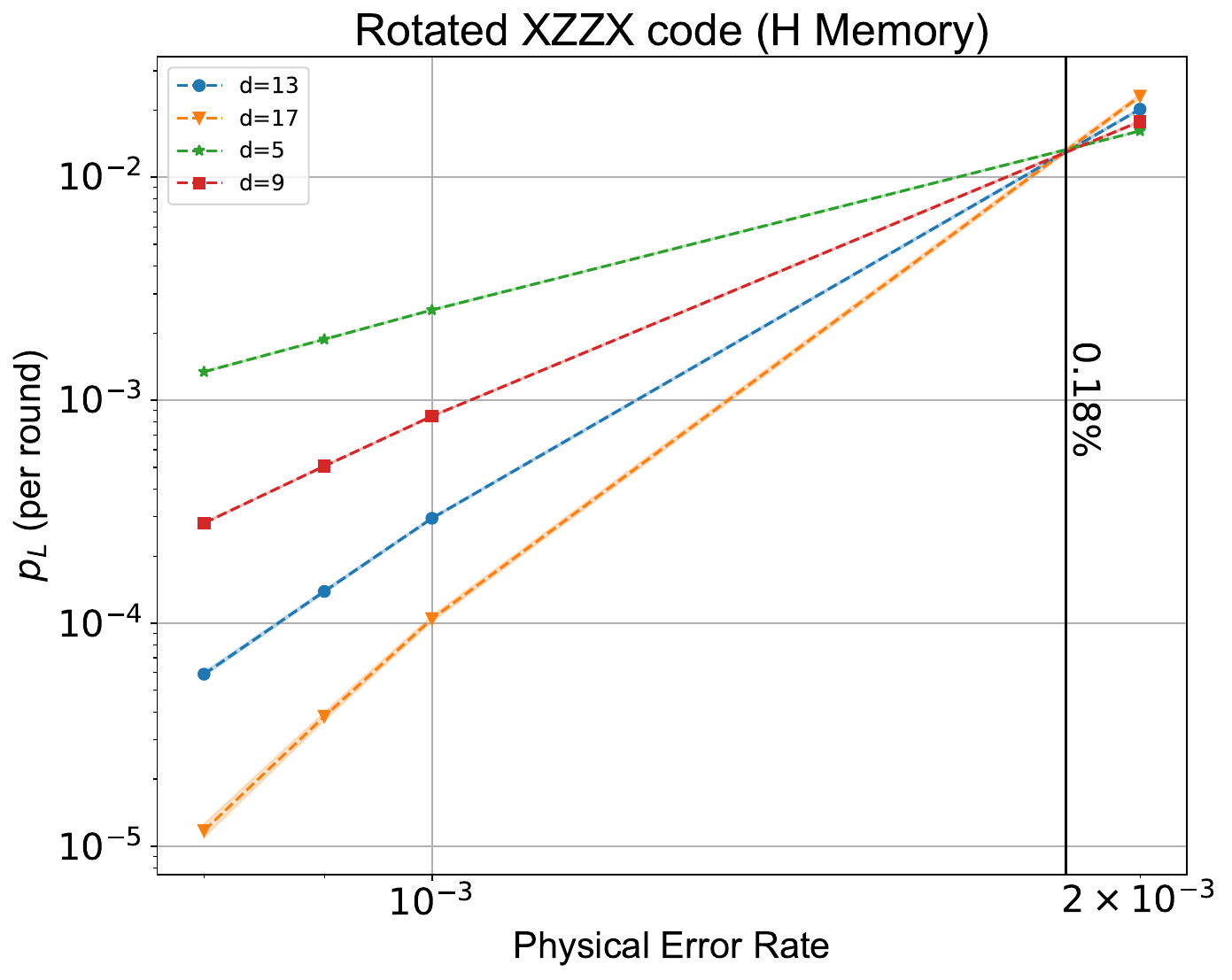}\label{fig:wat2}}
\caption{Logical error rate vs physical error rate $p$ (of two-qubit gates) for an $(N_x,N_y)=(2,3)$ SpinHex device at noise bias $\eta=100$. (a,b) Lines (top to bottom) show vertical (V) and horizontal (H) memory (see methods for details) at surface code distance $d=5,9,13,17$. A vertical black line marks the threshold at $0.18\%$. 
}
\label{fig:XZZXwaterfall}
\end{figure*}

\subsection{Noise model}
\label{Sec:NoiseModel}

Next, we simulate the performance of the XZZX surface code using a circuit-level noise model tailored to the device characteristics of SpinHex.
Errors are modeled using Pauli channels of one- and two-qubits depending on the operation performed, that is, Pauli errors are applied stochastically with an error rate $p$. The overall error rate of a gate is the sum of the probability of each individual Pauli error.
When all Pauli errors occur with equal probability, the noise is called depolarizing noise. 

For a single spin qubit, the probability of experiencing a bit-flip ($X$ error) or a bit-phase-flip ($Y$ error) is usually equal $p_x \approx p_y$. Meanwhile, the probability of experiencing a phase flip ($Z$ error) $p_z$ is higher, as given by the bias value,
\begin{align}
    \eta = \frac{p_z}{p_x + p_y} \sim \frac{T_1}{T_2},
\end{align}
where $T_1$ and $T_2$ refer to the relaxation and dephasing times \cite{approximatingdecoherence}.
In our circuit-level noise model, we model idling operations of the stabilizer circuit with biased noise.
For all other operations (H, CX, CZ, and SWAPs), we assume an implementation that is not bias-preserving.
Hence, we model those operations using depolarizing noise.
To account for typical spin-qubit fidelities, we assume the identical error rate $p$ for all two-qubit operations. The entangling gate error rate will serve as the baseline error rate, and we quantify the rest of error rates as a function of such value. Therefore, whenever we comment on device thresholds and error rates, those refer to the error rates of two-qubit operations and the rest are scaled as a function of those.
Recall that every CZ (or CX) operation of the stabilizer circuit requires implementing $[4(N_x+N_y)-10]$ additional SWAP gates.
Noise in single-qubit gates is modeled using a smaller gate error rate $p/10$.
Noise in state preparation and readout is modeled using higher error rates of $2p$ and $2p$, respectively.
Finally, to simulate the performance of the architecture we use the \textit{pymatching} implementation of the minimum-weight perfect matching decoder (MWPM) \cite{pymatching,sparseBlossom}.
For further details of the noise model and simulation, see Methods.

\subsection{Threshold and noise performance}

Figure \ref{fig:XZZXwaterfall} presents the simulated performance of the XZZX code for an $(N_x,N_y)=(2,3)$ SpinHex device.
The logical error rate is presented as a function of the physical error rate $p$ (of the two-qubit gate) for code distances $d=5,9,13,17$ (top to bottom), respectively.
A black line marks the threshold at a physical error rate of $0.18\%$.
Although data is reported for a noise bias $\eta=100$, it is worth noting that the threshold of the XZZX code remains constant for all bias values.
The threshold values for the rotated CSS surface code are similar.
(See the Supplementary Information for details.)
The consistent performance of the XZZX code with respect to the noise bias $\eta$ of the idling operations was the reason for choosing it.
In addition, note that the reported threshold of $0.18\%$ is lower than the $0.565\%$ threshold of the CSS and XZZX codes in the presence of standard depolarizing circuit-level noise as a result of the extra errors introduced by the SWAP gates \cite{rotvsunrot}.

The threshold value of $0.18\%$ implies that implementing the XZZX code on a (2,3) SpinHex device requires fidelity better than $99.82\%$ for two-qubit gates, $99.982\%$ for single-qubit gates, and $98.64\%$ for state preparation and measurement.
Such high fidelity values have yet to be achieved for spin-qubits (cf.~Tab.~\ref{tab:fidelity}). As such, they should be understood as target values for future implementations.

\begin{figure*}[]
\centering
\sidesubfloat[]{\includegraphics[width=7.9cm]{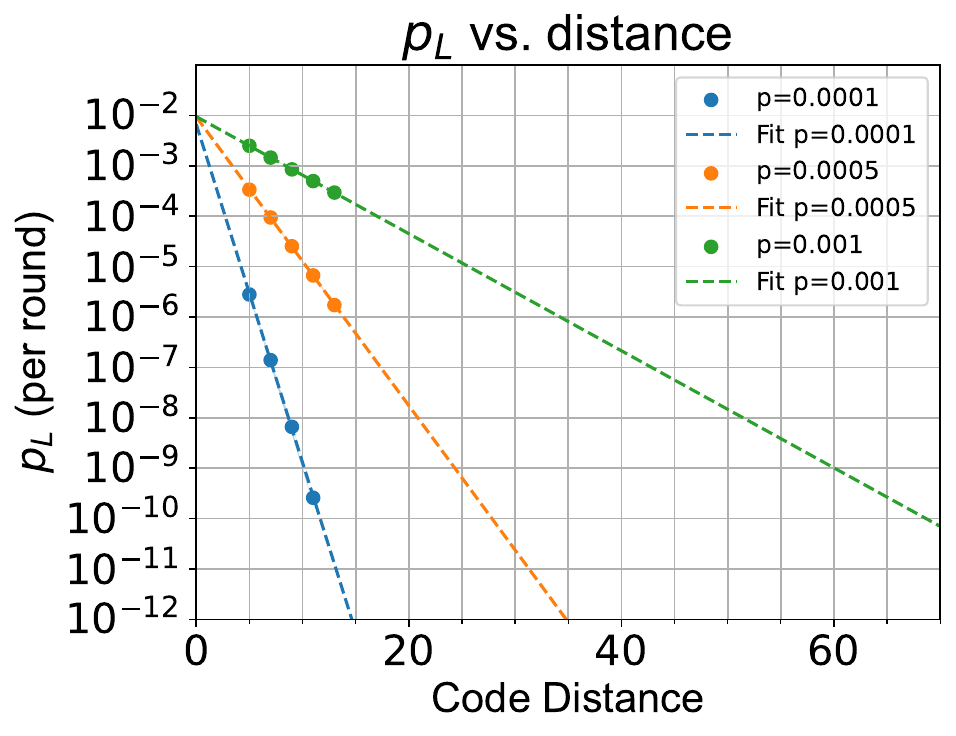}\label{fig:sub1}}\quad%
\sidesubfloat[]{\includegraphics[width=7.9cm]{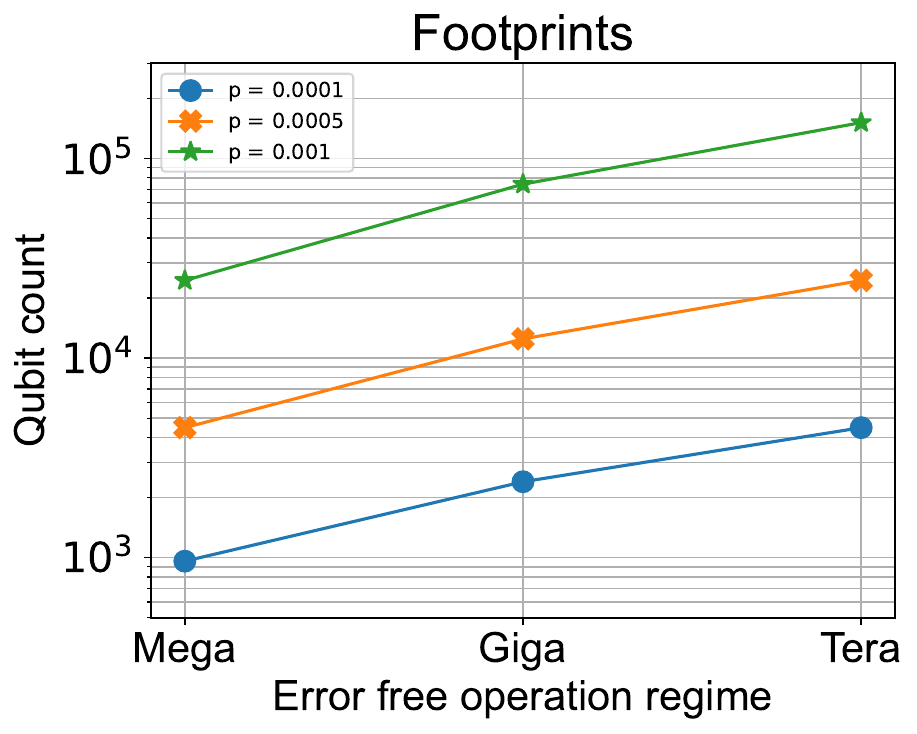}\label{fig:sub2}}
\caption{Code distances and number of physical qubits required to reach the Mega-, Giga- and TeraQuOp regimes in a (2,3) SpinHex device. (a) Logical error probability $p_L$ vs code distance $d$ for error rates $p=0.0001, 0.0005, 0.001$ (bottom to top). Dots represent numerically obtained values. Lines extrapolate the data to larger code distance. (b) Number of physical qubits required to reach logical error rates in the Mega-, Giga- and TeraQuOp regimes, using code distances from (a) with Eq.~\eqref{eq:QubitCount} for error rates $p=0.0001, 0.0005, 0.001$ (bottom to top).}
\label{fig:distproject}
\end{figure*}

\subsection{Qubit count and footprint}

\begin{figure}[b]
\centering
\includegraphics[width=0.85\textwidth]{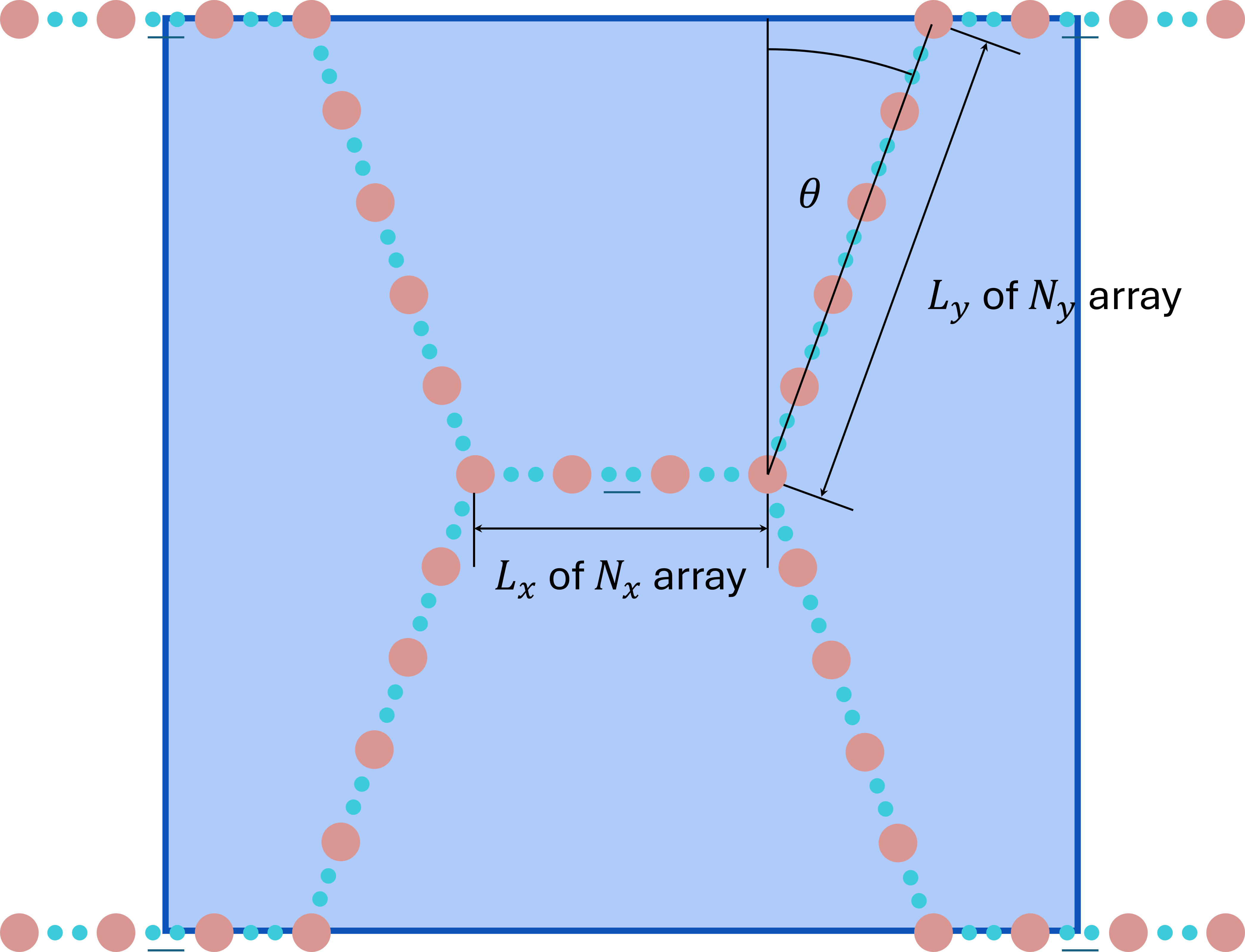}
\caption{
Schematic representation of a measurement square of an $(N_x,N_y)=(4,5)$ SpinHex device. Meaning of the symbols same as in Fig.~\ref{fig:HardwareGraph}.
}
\label{fig:MeasurementSquare}
\end{figure}

Next, we estimate the qubit count and footprint of a SpinHex device. To begin with, we count the number of couplers and double dots in a typical measurement square of an $(N_x,N_y)$ SpinHex device, see Fig.~\ref{fig:MeasurementSquare}. Assuming periodic boundary conditions, the number of couplers and double dots is given as
\begin{align}
    n_c    &= 2 N_x + 4 (N_y-2), \\
    n_{dd} &= 2(N_x-1)+4(N_y-1).
\end{align}
Further, assuming qubit quantum dots with a diameter (pitch) of 100 nm and couplers with a diameter of 500 nm, we estimate the length of horizontal and diagonal arrays as
\begin{align}
L_x &= 2\times100\mathrm{nm} (N_x-1) + 500\mathrm{nm} (N_x-1), \\
L_y &= 2\times100\mathrm{nm} (N_y-1) + 500\mathrm{nm} (N_y-1).
\end{align}
This length allows for upper bounding the area of each measurement square, Fig.~\ref{fig:MeasurementSquare}, consisting of 4 trapezoids as
\begin{align}
a_{s} = 4 \left(L_x+L_y\sin(\theta)\right)L_y\cos(\theta) \lesssim 4 \left(L_x+L_y\right)L_y.
\end{align}
Given that $d^2-1$ measurement squares host a logical qubit, we estimate the corresponding number of couplers, physical qubits and chip area per logical qubit as 
\begin{align}
    N_c &\approx [2 N_x + 4 (N_y-2)] (d^2-1),\\
    N_q &\approx 2[2(N_x-1)+4(N_y-1)] (d^2-1),\label{eq:QubitCount}\\
    A_q &\approx 1.96 \mu\mathrm{m}^2 (N_x+N_y-2)(N_y-1)(d^2-1).\label{eq:ChipSize}
\end{align}

\begin{table}[b]
 \centering
\caption{Code distance required to reach the MegaQuop, GigaQuop, and TeraQuop regimes.}
\label{tab:distances}
\begin{tabular}{ |c|c|c|c| }
 \hline
 Error rate & MegaQuop & GigaQuop & TeraQuop \\ 
 \hline
 $p=0.0001$ & $d=7$ & $d=11$ & $d=15$ \\ 
 \hline
 $p=0.0005$ & $d=15$ & $d=25$ & $d=35$ \\ 
 \hline
 $p=0.001$ & $d=35$ & $d=61$ & $d=85$ \\ 
 \hline
\end{tabular}
\end{table}

Next, we calculate the code distance $d$ and the number of physical qubits required to reach the MegaQuop, GigaQuop, and TeraQuop regimes, where a single logical qubit can perform a million, a billion, or a trillion error-free quantum operations, respectively. The results for an $(N_x, N_y)=(2,3)$ SpinHex device are shown in Fig.~\ref{fig:distproject}.

In Fig.~\ref{fig:distproject}(a), we plot the logical error rate as a function of code distance $d$ for three values of the error rate $p$. Lines denote exponential fits, which we use to extrapolate the code distance to the MegaQuop, GigaQuop, and TeraQuop regimes. The extracted code distance as a function of gate error rate are summarized in Tab.~\ref{tab:distances}.

In Figure \ref{fig:distproject}B, we use the extracted code distance together with Eq.~\eqref{eq:QubitCount} for an $(N_x,N_y)=(2,3)$ SpinHex device. 
For error rates $p=0.001$ qubit counts are prohibitively large. Even the MegaQuop regime requires approximately $24480$ physical qubits per logical qubit.
For error rates $p=0.0005$, it becomes feasible to reach the Megaquop regime with $4480$ physical qubits per logical qubit.
Finally, for error rates as low as $p=0.0001$, the same physical qubit footprint could allow us to reach error rates close to the desired TeraQuop regime.

Finally, we estimate the chip area for an $(N_x,N_y)=(2,3)$ SpinHex device.
We assume a device with (two-qubit gate) error rate $p=0.0001$.
This device allows a distance $d=15$ XZZX code to operate a single logical qubit with approximately $4480$ physical qubits in the TeraQuop regime.
Furthermore, assuming 10,000 logical qubits and using Eq.~\eqref{eq:ChipSize}, we estimate a SpinHex chip size of approximately 0.26cm$^2$.
Since a fault-tolerant processor would also require magic state factories and entangling operations, we estimate a $10$ times larger chip area of $2.6\mathrm{cm}^2$.

\subsection{Threshold versus area of the connection island}

\begin{figure}[]
\center
\includegraphics[width=\textwidth]{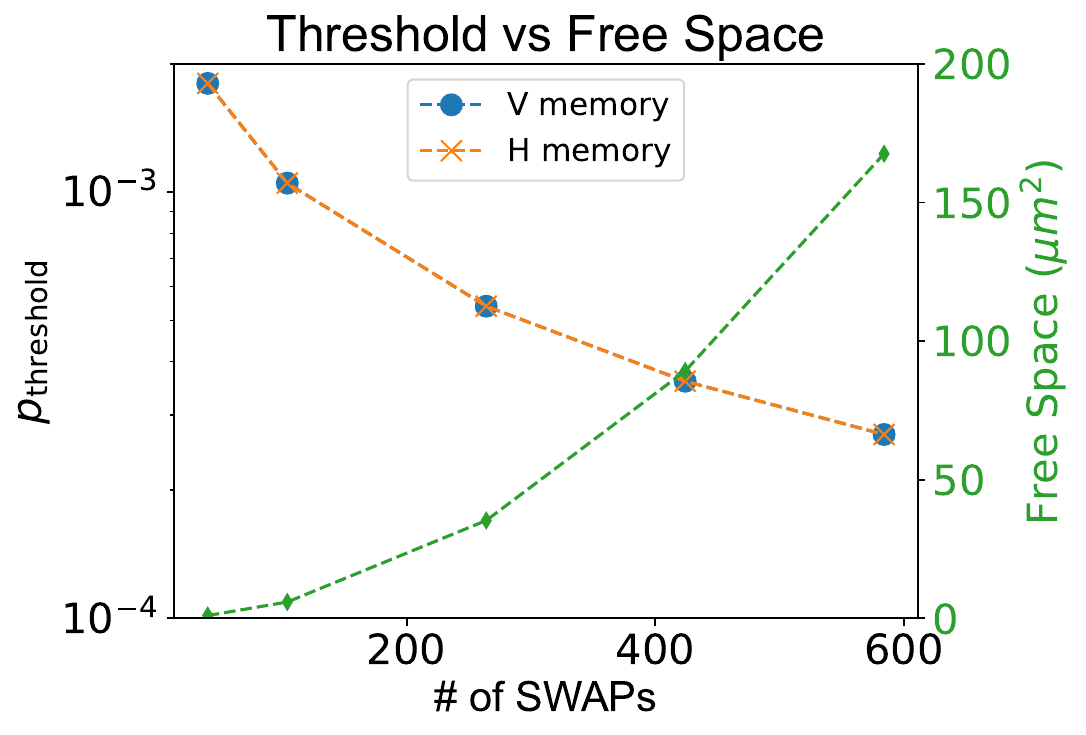}
\caption{Thresholds (left axis) and area of the connection island (right axis) vs the number of SWAPs per stabilizer measurement (cf.~Fig.~\ref{fig:circuit}) of a SpinHex device (yellow, blue and green curves), respectively.
}
\label{fig:tradeoff}
\end{figure}

One merit of the SpinHex device is the ability to integrate control electronics in the connection islands.
Such islands have an approximate area given as
\begin{align}
 A_{c} \approx L_x L_y = 0.49\mu\mathrm{m}^2(N_x-1)(N_y-1).
\end{align}
See Fig.~\ref{fig:DeviceDesign}(a) and Fig.~\ref{fig:MeasurementSquare}.
So far, our numerical calculations have focused on the SpinHex device $(N_x,N_y)=(2,3)$. 
However, larger connection islands with $N_x>2$ and $N_y>3$ may be required to host control electronics \cite{van2019impact, vandersypen2017interfacing, reilly2015engineering, boter2022spiderweb}.
Such islands would increase the number of SWAP operations required to implement the stabilizer circuit (cf.~Fig.~\ref{fig:circuit}) as
\begin{align}
   N_{SWAP} = 4\times2\times\left[2N_x-1 + 2(N_y-2)\right].
\end{align}
The increased number of SWAP operations would then lower the surface code threshold.
This trade-off is illustrated in Fig.~\ref{fig:tradeoff}.
The plot shows both the threshold and the area of the connection island as a function of $N_{SWAP}$ for SpinHex devices with $(N_x,N_y)=(2,3), (4,5), (9,10), (14,15)$ and $(19.20)$, respectively.
As can be seen, the code threshold decreases from $0.18\%$ to $0.027\%$ when almost $600$ SWAP operations are required. Meanwhile, the area of the connection island increases to $\sim150\mu m^2$.
We note that the realization of a device with such large connection islands is still beyond the experimentally achieved gate fidelity (cf.~Tab.~\ref{tab:fidelity}).

\section{Discussion}

\textbf{Summary:\ }
In this work, we introduce SpinHex, a semi-realistic spin-qubit device for operating surface codes.
SpinHex provides a 2D arrangement of spin qubits in a hexagonal lattice. Its core unit is a spinless quantum dot occupied by multiple electrons or holes. Such quantum dots act as couplers, which mediate voltage-controlled Heisenberg exchange \cite{mehl2014two, srinivasa2015tunable, Malinowski2019fast} between spin qubits in 2D at low capacitive crosstalk.
To evaluate the feasibility of SpinHex, we provide extensive numerical simulations. These simulations probe the performance of the rotated XZZX surface code under device-specific circuit-level noise.
For a minimal (2,3) SpinHex device, we report a threshold of 0.18\%. Further assuming an error rate of 0.01\%, we show that approximately $4480$ physical qubits can realize a logical qubit in the TeraQuop regime. (In this regime a logical qubit can execute a trillion error-free operations.) We further show that a SpinHex device operating $10,000$ logical qubits in the TeraQuop regime, can be hosted on a chip area of $2.6\mathrm{cm}^2$. A chip of this size can fit in a dilution refrigerator, and from that point of view, represents a feasible proposal for implementing surface codes with spin qubits.

\textbf{Benefits:\ }
We now list potential benefits of SpinHex:\
(i) Low capacitive cross-talk between spin-qubits, in particular, across junction elements.
(ii) Large distance between data and measurement qubits, potentially exceeding the correlation length of the noise environment.
(iii) Potential space for the collocation of control electronics side-by-side with the qubit arrays.

\textbf{Assumptions:\ }
We now list a couple of assumptions, omissions, and open questions, which need to be addressed in future work.
(a) SpinHex focuses on depletion gates. To make it work for SiMOS or Si/SiGe platforms, further modifications are needed. 
(b) For SiMOS or Si/SiGe platforms SpinHex requires the addition of micromagnets or microwave antennas for spin control.
(c) Multi-electron couplers have been verified in 1D \cite{mehl2014two, srinivasa2015tunable, Malinowski2019fast}. An experimental verification of 2D operation remains as future work. Verifying the operation with holes instead of electrons is also future work.
(d) So far, we focus on surface code operation. Meanwhile, universal quantum computing will require additional functionality to entangle logical qubits and perform magic state injection.
(e) SpinHex requires very high fidelity to operate efficiently ($99.99\%$ for SWAP, CX and CZ; $99.999\%$ for Hadamard; and $99.98\%$ for SPAM). Achieving these fidelity values is future work.

\section{Other}

\textbf{Code availability:} The code supporting the findings of this study can be found on GitHub: \url{https://github.com/pschnabl/spin-qubit-MEC-surface-code}.

\textbf{Competing interests:}
Hitachi co-authors filed the patent EP3975072A1.

\textbf{Acknowledgements:}
The authors thank the members of the Hitachi-Cambridge Laboratory and the Tecnun Quantum Information Group for their support and many useful discussions. This work was partially supported by the Spanish Ministry of
Economy and Competitiveness through the MADDIE project (grant No. PID2022-137099NBC44). J.E.M. is funded by the Spanish Ministry of Science, Innovation and Universities through a Jose Castillejo mobility grant for his stay at the University of Cambridge.

\textbf{Author contribution:}
R.M.O and F.M invented the architecture with input from C.G.S and N.M.
R.M.O designed the implementation of the surface code.
J.E.M., P.S and R.M.O designed the circuit-level-noise model for the XZZX code.
J.E.M and P.S performed the numerical simulations for the performance of the XZZX code and the associated footprint.
All authors contributed to the interpretation of the results and discussed the implications.
R.M.O., J.E.M. and N.M wrote the paper with input from all the co-authors.

\section{Methods}

\subsection{Circuit-level noise model}
\label{meth:cln}

\begin{table}[b]
\centering
\begin{tabular}{|c|c|c|c|c|}
 \hline
Operation & SiMOS(n) & Si/SiGe(n) & Ge/SiGe(p) & SiMOS(p) \\
\hline
Init    & $>99\%$ \cite{Huang2024}
        & --
        & --
        & --       \\
\hline
H       & $99.9\%$ \cite{Stuyck2024} 
        & $99.7\%$ \cite{Philips2022}
        & $99.97\%$ \cite{Wang2024}
        & $99.5\%$ \cite{Bassi2024}       \\
\hline
CZ/CX   & $>98.4\%$ \cite{Tanttu2024} 
        & $>99\%$ \cite{Xue2022}
        & $99.3\%$ \cite{Wang2024}                  
        & $--$ \cite{Geyer2024}\\
\hline
SWAP    &  --     
        &  --
        &  -- 
        &  --     \\
\hline
coupler &  --      
        &  --       
        &  --
        &  --      \\
\hline
Read    &  $99.2\%$ \cite{Oakes23} 
        &  $99\%$ \cite{Takeda2024} 
        &  $90\%$ \cite{riggelen2024coherent}
        &  low \cite{Bassi2024}   \\
\hline
Idle    &  --    
        &  --
        &  --
        &  --      \\
\hline
$T_1$   & $1-10^3$ms
        & $1-10^3$ms
        & $>\!\!1$ms \cite{Hendrickx2021}
        & $>\!\!10 \mu\mathrm{s}$ \cite{Camenzind2022}       \\
 \hline
$T_2^*$ &  $10-10^2 \mu s$
        &  $10-10^2 \mu s$
        &  $<\!\!400$ns \cite{Hendrickx2021}
        &  $<\!\!200$ns \cite{Camenzind2022}\\
 \hline
Bias    & $>10$
        & $>10$
        & $>2500$
        & $>50$   \\
 \hline
\end{tabular}
\caption{Experimentally reported fidelity values, see \cite{Stano2022}.}
\label{tab:fidelity}
\end{table}

To customize our circuit-level noise model, we assume that errors for all operations of the stabilizer circuit (CZ, CX, SWAPs, H, initialization, idling, and readout) occur independently.
To customize their relative strength, we consider the experimentally reported fidelity values in Tab.~\ref{tab:fidelity}.
Based on these values, we set the relative error probability as summarized in Tab.~\ref{tab:clnarchi}.

The gate error of two-qubit gates (CZ, CX) is modeled using two-qubit depolarizing noise with an error rate $p$.
Since single-qubit gates have an order of magnitude lower infidelity, we set the error rate of the Hadamard operation to $p/10$. Here, we model the noise using a single-qubit depolarizing channel.
Although state preparation and readout can have fidelity in excess of $99\%$, both can be challenging in experiments. Hence, we model both processes with an error rate of $2p$.

\begin{table}[b]
 \centering
\caption{Tailored circuit-level noise model for the proposed architecture based on silicon spin qubits.}
\label{tab:clnarchi}
\begin{tabular}{ |c|c|c| }
 \hline
 Operation & Error & Rate \\ 
 \hline
 CNOT/CZ & Depolarizing: $\{X,Y,Z\}^{\otimes 2}\setminus \{I^{\otimes 2}\}$ & $p$ \\ 
 \hline
 Hadamard & Depolarizing: $\{X,Y,Z\}$ & $p/10$ \\ 
 \hline
  Reset $\ket{0}$,$\ket{+}$ & Instead resets to $\ket{1}$ ($\ket{-}$) & $2p$ \\ 
 \hline
  Measurement & Flip the obtained result & $2p$ \\ 
 \hline
  SWAP & \makecell{Depolarizing error: $\{X,Y,Z\}$ \\ on the data/measurement qubit.}  & $0.8p$ \\ 
 \hline
  Idling & \makecell{Biased Pauli: $\{X,Y,Z\}$ with \\ $p_x=p_y=\frac{p_{idl}}{2(1+\eta)}$ and $p_z=\frac{\eta p_{idl}}{(1+\eta)}$} & $\xi\left(p/10\right)$ \\ 
 \hline
\end{tabular}
\end{table}

Similar to two-qubit gates, we model the errors of SWAP operations using two-qubit depolarizing noise with an error rate $p$. (We acknowledge that the fidelity of a coupler SWAP will likely be lower in initial experiments.)
We further note that a SWAP operation will generally exchange a data (or measurement) qubit with an ancillary qubit. This implies that only 12 of the possible 15 Pauli-errors in the two-qubit depolarizing channel will affect the data (or measurement) qubit. Thus, we model the error during a SWAP operation as a single-qubit depolarizing channel with reduced error rate $12/15p=0.8p$.

To complete the noise model, we need to discuss noise experienced by idling qubits. Silicon spin qubits present very high coherence times \cite{woottonSpinSC, strikisSC, veldhorst2014}. Since the operation times in silicon spin-qubit systems are very fast, idling may be seen as a negligible source of error or at least few orders of magnitude less probable. However, in other qubit technologies \cite{idlingSuper, idlingDD}, it has been observed that idling noise is higher when operating adjacent qubits. Here, we model the idling noise using an error rate $p/10$. Furthermore, we adjust the error rate according to the time it takes to execute an operation, using a scaling factor $\xi$. For example, idle qubits in the readout stage will experience errors at a rate $7(p/10)$, while idling qubits at a single-qubit gate step will experience an error rate of $1/10(p/10)$. The baseline $\xi=1$ is set when executing a CX, CZ, or SWAP gate on other qubits. Finally, silicon spin qubits present a strong bias towards dephasing errors and, thus, the idling steps will be modeled by means of a Pauli channel with a bias $\eta$.
A summary of our model is given in Tab.~\ref{tab:clnarchi}.

\subsection{Numerical simulations}

Extensive Monte Carlo simulations of the CSS and XZZX codes have been performed in order to estimate the performances of the codes considering the circuit-level noise model described before. We implement the noisy extraction circuits in order to perform the sampling of the errors by using Stim \cite{stim}. Stim considers the check measurements upon a set of syndrome extractions altogether with a final measurement of the data qubits. In order to conduct the memory experiments of the CSS (See Supplementary Material) and XZZX rotated surface codes, two different memory experiments are conducted depending on which initial state is aimed to be preserved:
\begin{itemize}
    \item CSS surface code: $Z$ memory experiment in which all data qubits are initialized to the $\ket{0}$ state and finally measured in the $Z$ basis; and the $X$ memory experiment in which all data qubits are initialized to the $\ket{+}$ state and measured in the $X$ basis.
    \item XZZX surface code: the horizontal (H) memory experiment in which data qubits are initialized in an interchanging pattern of $\ket{+}$ and $\ket{0}$ states starting from $\ket{+}$ for the top left data qubit and finally measured in their corresponding bases; and the vertical (V) memory experiment in which data qubits are initialized in an interchanging pattern of $\ket{0}$ and $\ket{+}$ states starting from $\ket{0}$ for the top left data qubit and finally measured in the corresponding basis. The horizontal and vertical memory names come from the direction of the strings forming logical operators in each of the cases.
\end{itemize}

The operational figure of merit we use to evaluate the performance of the codes is the logical error probability, $p_L$, per extraction round. We ran $3d$ rounds of syndrome extraction to reduce time-boundary edge effects coming from the fact that the first and last extraction rounds are less noisy than the ones in the bulk \cite{rotvsunrot}. Once the circuits are run to collect the samples, we use the \textit{pymatching} implementation the MWPM decoder in order to aim for recovery and determine, if a logical error has occurred or not \cite{decoders,pymatching,sparseBlossom,bpotf_2024}. To decode at the circuit-level, one must get the detector error model, which we do using Stim \cite{stim,bpotf_2024}. In order to have good enough statistical accuracy from the Monte Carlo simulation estimations, we began collecting samples with a ceiling of twenty million circuit shots and a hundred thousand logical errors. For the lowest logical error probability points, i.e. $d=13,15$ at $p=0.0001$, we increased the shot ceiling up to ten billion shots. We highlight regions showing $p_L$ values for which the conditional probabilities $P(p_L|k)$ are within a factor of $1000$ of the maximum likelihood estimation, $p_L = k/n$, assuming a binomial distribution, only for the numerically estimated points, but are too small to be observed \cite{teraHoneycomb}.


%

\end{document}



\title{Supplementary Material for 2D Spin-Qubit Architecture with Multi-Electron Couplers for Surface Error Correcting Code}

\author{Rubén M. Otxoa}
\affiliation{Hitachi Cambridge Laboratory, J. J. Thomson Avenue, Cambridge CB3 0HE, United Kingdom.}
\author{Josu Etxezarreta Martinez}
\affiliation{Department of Basic Sciences, Tecnun - University of Navarra, 20018 San Sebastian, Spain.}
\affiliation{Cavendish Laboratory, Department of Physics, University of Cambridge, Cambridge CB3 0HE, UK.}
\author{Paul Schnabl}
\affiliation{Department of Basic Sciences, Tecnun - University of Navarra, 20018 San Sebastian, Spain.}
\author{Normann Mertig}
\affiliation{Hitachi Cambridge Laboratory, J. J. Thomson Avenue, Cambridge CB3 0HE, United Kingdom.}
\author{Charles Smith}
\affiliation{Hitachi Cambridge Laboratory, J. J. Thomson Avenue, Cambridge CB3 0HE, United Kingdom.}
\author{Frederico Martins}
\affiliation{Hitachi Cambridge Laboratory, J. J. Thomson Avenue, Cambridge CB3 0HE, United Kingdom.}

\maketitle

\section*{Surface codes: the rotated planar CSS and the XZZX codes}

Surface codes: the rotated planar CSS and the XZZX codes
Surface codes stand as some of the most promising platforms for fault-tolerant quantum computing as a result of their locality of interactions (parity checks) and their high tolerance to quantum noise \cite{decoders,fowlerSC}. Furthermore, transversal operations and magic state injection are very well understood in surface codes implying that their use for fault-tolerant quantum algorithm implementation is very appropriate \cite{fowlerSC,gidneyShor}.

The rotated planar code (we will also refer to it as the CSS surface code) is one of the most well-known surface codes, encoding a single logical qubit by means of $d^2$ physical qubits with minimum distance $d$ \cite{decoders}. This code is a result of optimizing the planar code, which comes from cutting the periodic conditions of Kitaev's toric code in order to have a strictly planar qubit layout \cite{kitaev}. Experimental realizations of the rotated planar code have been realized in several qubit platforms, e.g. superconducting and neutral atom hardware \cite{googleSurf,bluvstein2023logical,google24}. The stabilisers of the rotated planar code are four body interactions defined as $X_aX_bX_cX_d$ and $Z_aZ_bZ_cZ_d$, where $a,b,c,d$ refer to qubits that are nearest neighbors \cite{fowlerSC}. Stabiliser measurements (SM) allow us to extract the so-called syndrome which indicates whether an errors has occurred \cite{decoders,QECbeginners}. In such extraction circuits, the data qubits interact with the checks (or measurement) qubits by means of two-qubit gates (CNOT, CZ), see Supplementary figure \ref{fig:circuitCSS}. Note that a class of check is susceptible to errors that anti-commute with it, which is why the $Z$-checks detect $X$ errors and vice versa. Thus, this code belongs to the Calderbank-Shor-Steane class (CSS). Finally, the logical operators of this code are strings of $Z$ (logical $Z$) or $X$ (logical $X$) operators that go from boundary to boundary, top to bottom for $Z$ and left to right for $X$

Alternatively, other sets of stabilisers can be used leading to other variants of the rotated planar code. Generally, surface codes admit a technique named as Clifford deformation, where the original stabilisers are modified by means of single-qubit Clifford operations \cite{cliffordDeformation}. The resulting codes may be topologically equivalent to the original code, but they may exhibit better capabilities for certain noise models. Bias towards dephasing has shown to be beneficial for QEC \cite{decoders,toninid,beliefmatching,rectBias,biasQLDPC,recursiveMWPM,tuckettBiasSC,xzzx,tailoredXZZX,tuckettxy}. This is the case of the XZZX code, obtained by applying a Hadamard gate to every other qubit of the CSS rotated surface code \cite{cliffordDeformation,xzzx}. The XZZX code shows excellent capabilities to deal with noise that is strongly biased towards dephasing \cite{xzzx}. The resulting stabilisers have all the form of $X_aZ_bZ_cX_d$, where the labels are nearest neighbor qubits, as before. Furthermore, the logical stabilisers do also become into chains of interchanging $X$ and $Z$ operators. Importantly, there is a single logical operator purely consisted of $Z$ operators, from the left top corner to the bootom right \cite{decoders}. Therefore, whenever noise is strongly biased towards dephasing, it is much less unlikely that such error operator occurs than for the CSS version. In addition, all check qubits are susceptible to $Z$ errors.

\begin{figure*}[!ht]
\centering
\includegraphics[width=16cm]{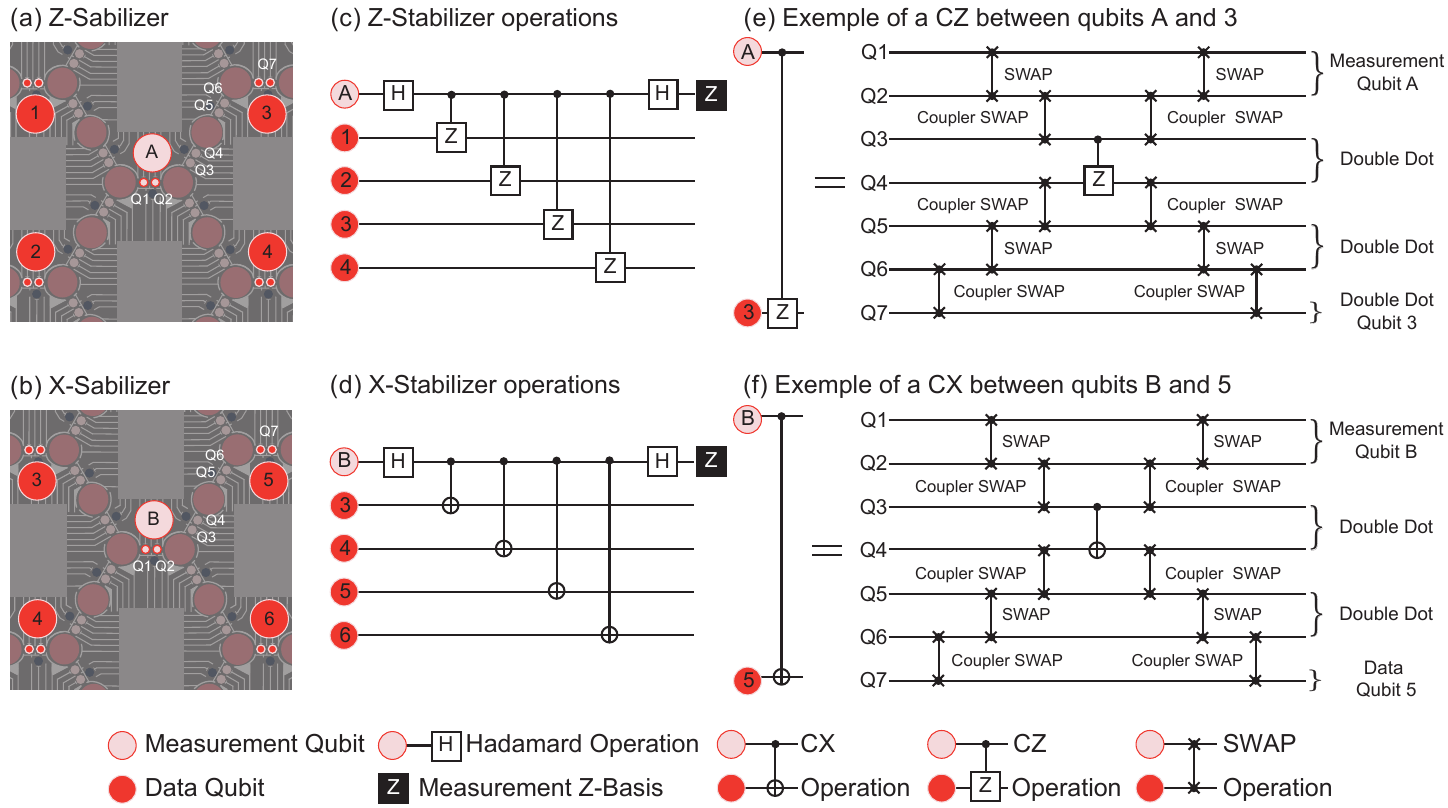}
\renewcommand{\figurename}{Supplementary Fig.}\setcounter{figure}{0}\caption{Illustration of X- and Z-stabiliser operation in the architecture proposed here. (a,b) Zoom into a Z- and X- Stabiliser of qubit units (1,2,3,4,A) and (3,4,5,6,B), respectively. Qubit units A and B are the data qubits of the stabiliser. (c,d) Quantum circuits to implement the Z- and X-stabilisers. (e,f) Illustration of an implementation of a CNOT gate between data qubit of qubit unit A (B) and measurement qubit of qubit unit 3 (or 5) for Z-stabiliser (or X-stabiliser) operation, respectively. To implement the CNOT gates one uses SWAP (within the qubit units and across MECs) along the chain of physical qubits labelled Q1 – Q7. 
}
\label{fig:circuitCSS}
\end{figure*}

\section*{Comparing the rotated CSS code and the XZZX variant}
As mentioned in the main text, we select the XZZX variant of the surface code to be implemented rather than the standard rotated CSS one. 

In Supplementary figure \ref{fig:cssVSxzzx} we plot the threshold values for both codes as a function of the bias on a $(2,3)$ SpinHex architecture. In order to get those plots, we computed more points than for the waterfall plot in the main text around the threshold value for precision. It can be seen that the threshold of both variants of the surface code are the same, i.e. $0.18\%$ irrespective of the value of the bias. The reason for this invariance with respect to the bias comes from the fact that the noise at the circuit-level is dominated by depolarizing noise See Methods in the main text, the only biased noise comes from idling), especially due to all the SWAP steps that add a significant amount of depolarizing noise. As a result, either choice should present a similar performance for the architecture proposed at this point. However, we have selected the XZZX code proposal because a strong bias towards dephasing is a characteristic of spin qubits over semiconductors (See Methods). In this sense, there is no proposal to leverage the noise bias for two-level system qubits such as the theoretical bias-preserving CNOT gates over cat qubits \cite{biaspreservingCNOTs}. Further studies may be able to exploit such a feature for semiconductor spin qubits and, thus, we have preferred to have a bias tailored code as the baseline.

\begin{figure*}[!ht]
\centering
\includegraphics[width=8cm]{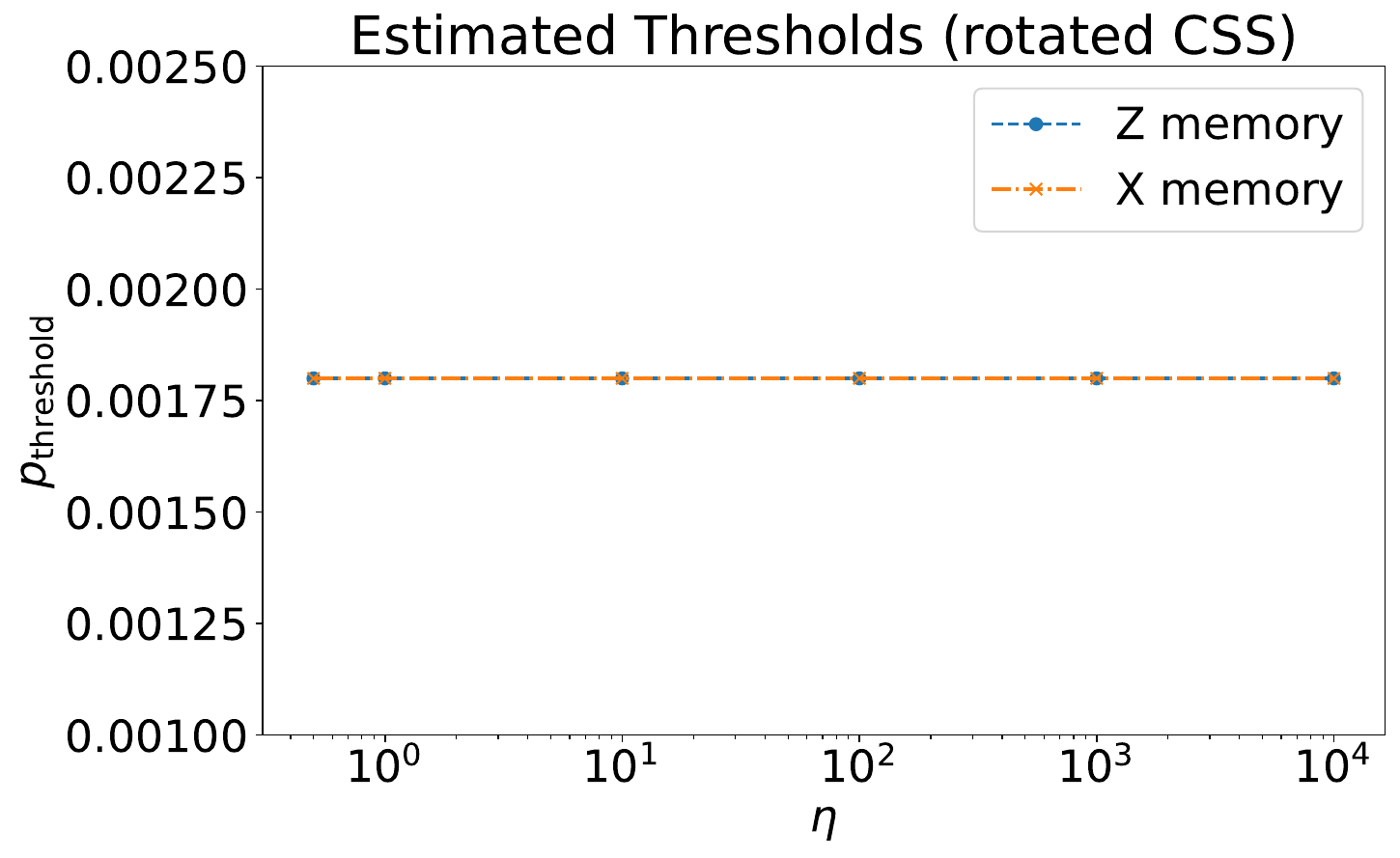}
\includegraphics[width=8cm]{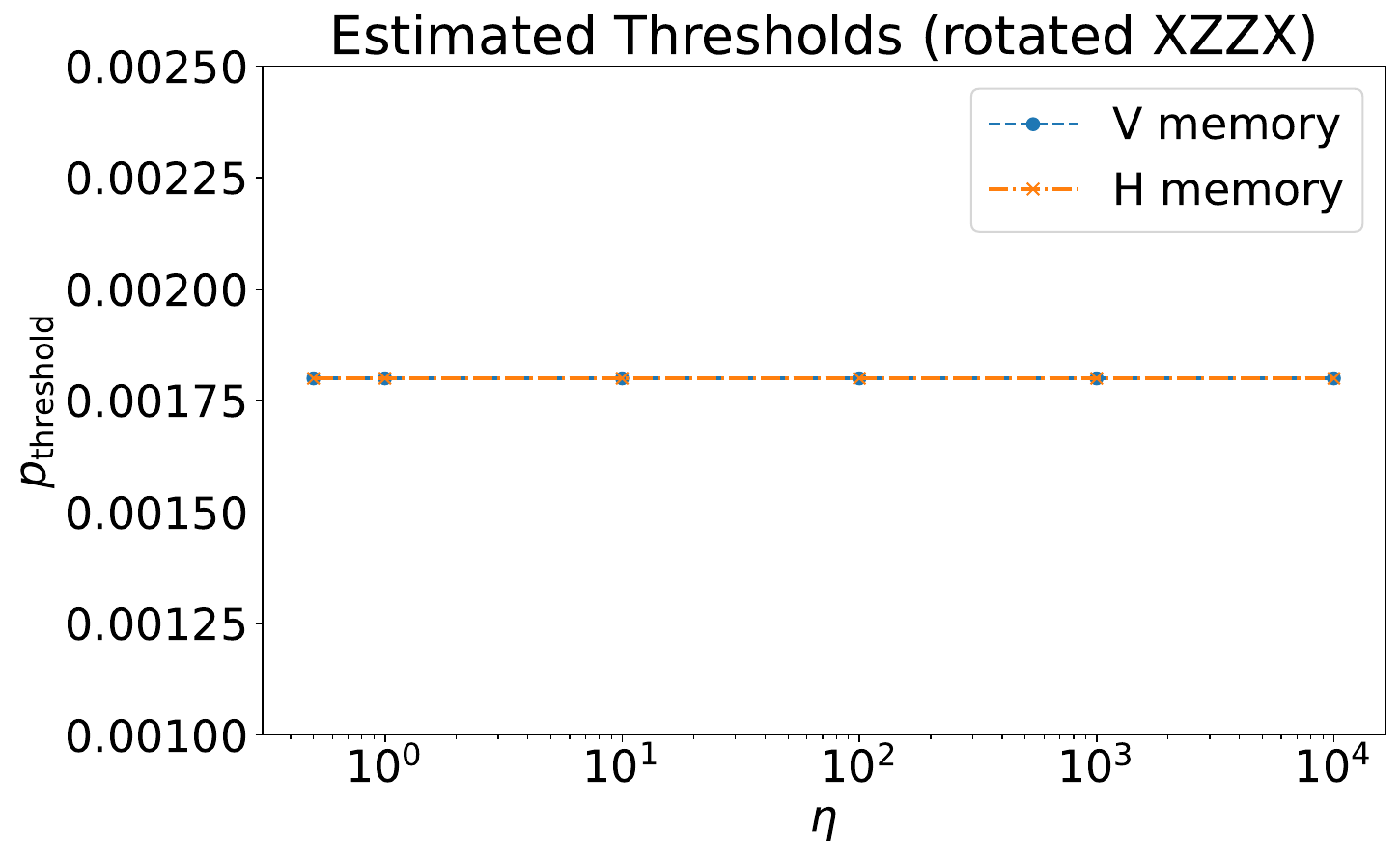}
\renewcommand{\figurename}{Supplementary Fig.}\setcounter{figure}{1}\caption{Thresholds of the rotated CSS and XZZX surface codes as a function of the bias in SpinHex $(2,3)$. We plot the X and Z memories for the rotated CSS code (left) and the H and V memories for the XZZX code (right). Both codes present the same threshold independent of the idling bias, i.e. $0.18\%$.}
\label{fig:cssVSxzzx}
\end{figure*}

%
